\documentclass[structabstract]{aa}
\usepackage{graphicx}
\usepackage{txfonts}
\begin{document}
   \title{Robust correlators}

   \author{P. Fridman
          \inst{}
          }

   \institute{ASTRON, Oudehoogeveensedijk 4,  Dwingeloo, 7991PD, The Netherlands\\
              \email{fridman@astron.nl}
             }

   \date{}



  \abstract
   {Radio frequency interference (RFI) already limits the sensitivity of existing radio telescopes in several frequency bands
   and may prove to be an even greater
   obstacle for  future generation instruments to overcome.}
   {I aim to create a structure of  radio astronomy correlators which will be statistically
   stable (robust) in the presence of interference.
   }
   {A statistical analysis  of the mixture of \emph{system noise + signal noise + RFI} is proposed here which could be  incorporated into the
   block diagram of a correlator. Order and rank statistics are especially useful when calculated
   in both temporal and frequency domains.}
   {Several new algorithms of  robust correlators are proposed and investigated here. Computer simulations
   and  processing of  real data demonstrate the efficacy of the proposed algorithms.}
   {}


   \maketitle
%

\section{Introduction}

  Correlators are central to the signal processing system of radio interferometer, \cite{thompson}.
  Signals received from radio sources mixed with system noise (sky noise  $+$ receiver noise)
  can be represented as stochastic processes  with a Gaussian (normal) distribution. Each pair of random numbers at the input of the correlator
is described as a bivariate random value $(X,Y$) with the distribution ${\cal N}(\mu _{X},\mu _{Y,}\sigma _{X}^{2},\sigma _{Y}^{2},\rho_{0} )$
where $\mu _{X}=\mu _{Y,}=0$ are the mean values, $\sigma _{X}^{2}$ and $\sigma _{Y}^{2}$ are  variances  proportional to the intensity of the input noise and $\rho$ is the correlation coefficient (normalized visibility for a particular baseline). From the sequences $x_{1},...,x_{N}$ and $y_{1},...,y_{N}$   the correlator   calculates the so-called Pearson's correlation coefficient
\begin{equation}
r_{P}=\frac{\sum_{i=1}^{n}(x_{i}-\overline{X})(y_{i}-\overline{Y})}{\sqrt{\sum_{i=1}^{n}(x_{i}-\overline{X})^{2}}\sqrt{\sum_{i=1}^{n}(y_{i}-\overline{Y})^{2}}},
\end{equation}
where $\overline{X}$ and $\overline{Y}$  are the arithmetic averages of $x$ and $y$, respectively.
The value $\lim_{n \rightarrow \infty}r_{P} =\rho_{0}$. This is valid both for XF and FX correlators
(the correlation of complex numbers is reduced to separate correlations of the real and imaginary components).\\
The sample correlation coefficient $r_{P}$ is {\bf very sensitive to outliers} (samples  which are not consistent with the normal distribution
${\cal N}(\mu _{X},\mu _{Y,}\sigma _{X}^{2},\sigma _{Y}^{2},\rho_{0} )$, .i.e., the estimate (1) is not statistically robust, (Gnanadesikan\cite{gnad97},
 Huber\cite{huber}).
 Radio frequency interference (RFI) can produce considerable outliers,  yielding a bias of  $r_{P}$ and increasing the standard deviation ({\em rms})
 of $r_{P}$. In this paper the terms ``RFI" and ``outliers" are  interchangeable.

  Methods  of robust statistics allow  stable estimates to be obtained in the  presence of outliers in several radio-astronomical applications
  (Fridman\cite{fridman}).
 Here these methods are applied to  correlation measurements of radio interferometers.

  There are different types of RFI (Fridman\&Baan\cite{fridmanbaan}) and they differ  significantly over the radio astronomy frequency band.
   Strong and impulse-like RFI on the time-frequency plane are visible in practically all frequency bands of LOFAR, and  these types of RFI  are
   considered here  because   data from the LOFAR  Core Station (CS1)  are used as examples  later in this paper.
   Fig. 1 give an impression of the  types of interferences in one sub-band: the auto-spectrum of the system noise + RFI  from the LOFAR CS1
    is represented in this figure with  high spectral and time resolution.

 \begin{figure}
 \includegraphics[width=8cm]{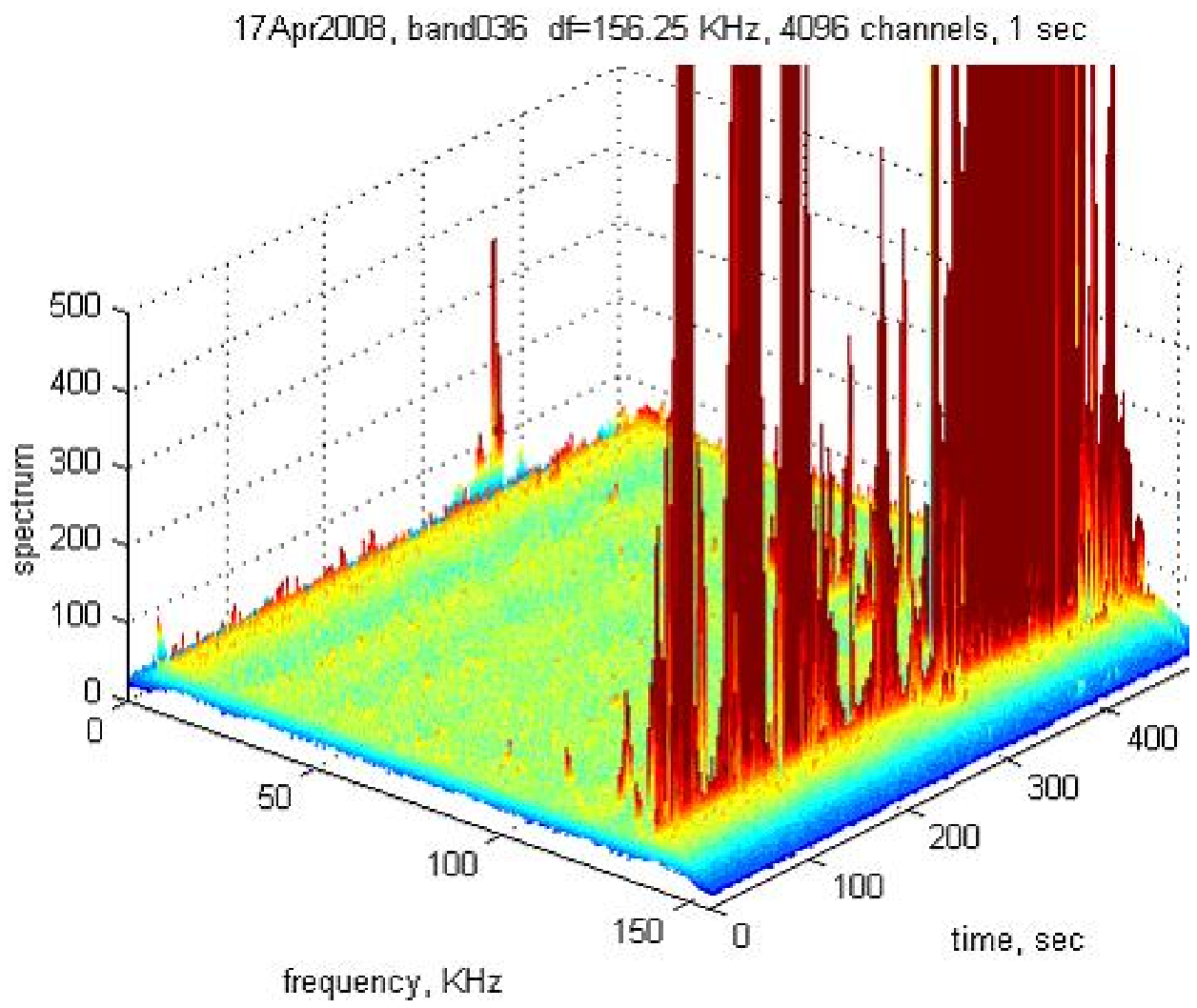}
 \caption{Autospectrum of  system noise + RFI from LOFAR CS1 at the central  frequency $f_{0}=217.1875MHz$, bandwidth $\Delta f=156.25KHz$,
 frequency resolution $\delta f\approx 38Hz$, integration time $t\approx 1s$.}
  \end{figure}

   The following approximate classification is proposed for  the strong RFI visible at LOFAR:\\
a) Narrow-band  and persistent RFI;\\
b) Narrow impulses ($<1sec, <100Hz$) on the time-frequency plane;\\
c) Impulse-like RFI in the time domain (several seconds) which are also  wide-band (5 -10 MHz);\\
d) Random-form bursts on the  time-frequency plane.\\
There are several types of RFI which cannot be treated as outliers: they may be  persistent and wide-band, weak or strong, fully
or partly correlated at the sites of  a radio interferometer. Methods  using  the spatial separation of the source of RFI and
the radio source  can effectively address this issue, see (Ellingson\&Hampson\cite{ellingson}, Jeff et al.{\cite{jeff},
Kesteven\cite{kesteven}, Cornwell et al.\cite{corn}).

Several types of  {\em robust} correlators that are able to  mitigate the impulse-like, strong RFI in both temporal and frequency domains
 have been  studied in this paper.
They are used in applications where input data are contaminated with outliers, and these correlators are  statistically  more stable than  correlator  (1).
 Some of them could be used in radio astronomy, especially in  radio interferometric systems with  {\em software}  correlators
 where the calculation of visibilities is carried out on  general-purpose  computers,  as in  (LOFAR\cite{lofar} and JIVE, \cite{kruithof}).
 Software correlators are, by definition, much more
 flexible than  traditional hardware  correlators: any algorithm adapted or modified for a particular observation can be optionally downloaded as a subroutine.\\
 There are two operations in the numerator of (1): multiplication of the input samples of $X$ and $Y$ and summation (averaging).
Here it is proposed that they be modified to make the
 correlator more robust.\\
 The new features  can be   introduced in the first operation    to analyze the statistics of $X$ and $Y$ and
 to introduce a type of  editing in order to eliminate outliers.\\
 The second operation, summation, which is  considered as  post-correlation averaging,  can be divided into three steps:
  primary averaging over a time
 interval which is not too long  to smooth RFI bursts appearing at this stage above the noise level,\\
 RFI mitigation \\
 and secondary averaging  to the required time interval depending on the observational specifications
  (wavelength, baseline, radio source properties), see  end of  section 3 and Figs. 13  - 15.

Different types of correlators  described in the following section are compared  with  Pearson's correlator (1) using two criteria:\\
1. The {\em bias} of the estimate $\widehat{\rho}$ produced  by RFI compared to the input correlation coefficient $\rho_{0}$;\\
2. The effectiveness  of an estimator is judged by the {\em rms} at the output of the correlator in both the presence  and in the absence  of outliers.

Computer simulations  were performed to estimate these values, of the bias and {\em rms}. Also some results    of the processing of CS1 data will be shown.


\section{Estimators of the correlation coefficient}
There are two classic estimators of correlation coefficients using {\em ranks} of  samples instead of the samples themselves, (Kendall\cite{kendall}).\\
Let two input sample sequences $x_{1},...x_{n}$ and $y_{1},...y_{n}$ be sorted in ascending order: $x_{(1)}\leq x_{(2)}\leq ...\leq x_{(n)}$  and
$y_{(1)}\leq y_{(2)}\leq ...\leq y_{(n)}$. The $i$th value $x_{(i)}$ is called $i$th-{\em order statistic}. Each sample $x_{j}$ has its $k$th position
in the sorted series  $x_{(1)},...x_{(n)}$. This number $k$ is the {\em rank} of the  sample $x_{j}$ and is denoted by $p_{j}(=k)$. Similarly, the rank of
 $y_{j}$ is denoted by $q_{j}$. Let $(x_{i},y_{i})$ and $(x_{j},y_{j})$ with $i=1,...n$ and $j=i+1,...n$ be two data pairs from the original data sequences.
 If $p_{j}-p_{i}$ and $q_{j}-q_{i}$ have the same sign, these two data pairs are said to be {\em concordant}, otherwise, they are {\em discordant}.
 Let $n_{c}$ be the number of concordant pairs and $n_{d}$  the number of discordant pairs. It is clear that $n_{c}+n_{d}=n(n-1)/2$.
 \subsection{Spearman's correlator}
 The Spearman's correlation coefficient is calculated  by
 \begin{equation}
 r_{SP}=1-\frac{6\sum_{i=1}^{n}(p_{i}-q_{i})^2}{n(n^{2}-1)}.
 \end{equation}
 The bivariate correlation coefficient corresponding to  Pearson's $r_{P}$ can be restored using the relationship:
\begin{equation}
\widehat{\rho }=2\sin (\frac{1}{6}\pi r_{SP}).
 \end{equation}
 \subsection{Kendall's correlator}
 Kendall's correlation coefficient is calculated by
 \begin{equation}
 r_{KND}=\frac{2(n_{c}-n_{d})}{n(n-1)}.
 \end{equation}
The bivariate correlation coefficient corresponding to  Pearson's $r_{P}$ can be restored using the relationship:
 \begin{equation}
\widehat{\rho }=\sin (\frac{1}{2}\pi r_{KND}).
\end{equation}
 \subsection{Correlator using sums and differences }
 One of the first  constructions of  a robust correlator is based on the quarter square identity (Gnanadesikan\&Kettenring\cite{gnad}, Huber\cite{huber}):
 \begin{equation}
 cov(X,Y)=1/4[var(X+Y)-var(X-Y)],
 \end{equation}
 where $cov$ denotes covariance  and $var$ denotes variance.
 The correlation coefficient can be obtained with
\begin{equation}
r_{GK}=1/4\frac{var(X+Y)-var(X-Y)}{\sqrt{var(X)var(Y)}}.
\end{equation}

Therefore, any robust estimators of variance can be used for this correlator (Fridman\cite{fridman}). Several of them are  applied here to (7).

\subsubsection{Trimming}

 Samples $Z_{1}=X+Y$  and   $Z_{2}=X-Y$ are sorted in ascending order: $z_{(1)},...z_{(n)}$.
 Let $\gamma$ denote the chosen amount of trimming, $0 \le \gamma\le 0.5$ and $ k=[ \gamma n]$.
 Sample-trimmed variance is computed by removing $k$ of the largest and $k$ of the  smallest data and using the values that remain:
\begin{eqnarray}
var_{trim}=\frac{K_{trim}}{n-2k}\sum_{i=k}^{n-k}(z_{i}-\widehat{\mu }_{trim})^{2}\\
\widehat{\mu }_{trim}=\frac{1}{n-2k}\sum_{i=k}^{n-k}z_{i}, \nonumber
\end{eqnarray}
where $\mu_{trim}$ is the sample mean of the trimmed data. Trimming
lessens the variance of  data and the coefficient $K_{trim}$ makes
$var_{trim}$ a consistent estimator for  data with a normal
distribution.   The value of  $K_{trim}$ depends on  $\gamma$, (Fridman\cite{fridman}), but in the context of equation (7),
this is not important, because $K_{trim}$  appears symmetrically in the numerator and denominator of (7).
\subsubsection{Winsorization}
A sample $Z$ is sorted in ascending order.
 For the
chosen $0 \leq \gamma  \leq 0.5$ and $k=[\gamma  n]$, Winsorization
of the sorted data consists of setting
\begin{equation}
W_{i}=\left\{ \begin{array}{lcl}z_{(k+1)}, & if & z_{(i)}\leq z_{(k+1)}\\
                                z_{(i)},   & if  & z_{(k+1)}<z_{(i)}<z_{(n-k)}\\
                    z_{(n-k)}, & if & z_{(i)}\geq z_{(n-k})\\

  \end{array} \right.
\end{equation}

The Winsorized sample mean is $\widehat{
\mu_{w}}=\frac{1}{n}\sum_{i=1}^{n}W_{i}$ and the Winsorized sample
variance is
\begin{equation}
var_{wins}=\frac{1}{n-1}\sum_{i=1}^{n}(W_{i}-\widehat{\mu _{w}})^{2}.
\end{equation}
The essence of Winsorization is  to replace the $k$ of the lowest and $k$ of the highest samples of the sorted data $z$ by the values of their nearest neighbors
$z_{(k+1)}$ and $z_{(n-k)}$, respectively.

\subsubsection{Median Absolute Deviation (MAD)}
This estimate of the variance of    sorted data $z_{(1)} \leq z_{(2)}\leq...\leq
z_{(n)}$ is defined by
\begin{equation}
var_{med}=(1.483\times med_{1\leq i\leq n}\{\left|
z_{i}-med(z_{i})\right| \})^2,
\end{equation}
where
\begin{displaymath}
\left. \begin{array}{lll}
med= & 0.5(z_{(m)}+z_{(m+1)}), & n=2m,\\
med= & z_{(m+1)},& n=2m+1.\\

  \end{array} \right.
\end{displaymath}

 Using $med(z_{i})$ in the place of sample mean and $(med\left|z_{i}-med(z_{i})\right|)^2$ in the place of sample variance provides a more
robust estimate of variance:  only central samples of sorted data (central order statistics) are used, according to the definition of
the median, and  outliers  are excluded. This procedure works well with  data contaminated by up to  $\approx 50\%$  outliers , but the
{\it rms} of the estimator (11) is larger than that of the classic variance estimate (see section 3, Table 4).

\subsubsection{ $Qn$  estimate}
This estimate is proposed in \cite{rous93} and  is moderately effective. It combines the
ideas of the  Hodges-Lehmann estimate and the Gini estimate,
(\cite{kendal67}):
\begin{equation}
var_{Qn}=2.108\{\left| x_{i}-x_{j}\right| ,i< j\}_{(k)},
\end{equation}
where  $k={h \choose 2}$ and $h=[n/2]+1$, that is $ k \approx {n \choose 2}/4$.
 This estimator works as follows: the  interpoint (pairwise) distances $\left| x_{i}-x_{j}\right|, i<j,$
 are sorted in ascending order. The $kth$ value of this sorted sequence (the $kth$ order statistics) multipled by consistency factor
 2.108 is  then taken as the estimate of variance. The value of the consistency factor  is also not critical,  as for trimming (section 2.3.1).

Other robust algorithms can also be applied (for example, the M-estimator, (Huber\cite{huber}) but here
 I was not attempting to make a comprehensive study of all types of robust correlators.
Rather I wished to direct attention to the options hitherto unused
in  designing correlators.

\section{Testing  the algorithms}
\subsection{Mitigation of outliers in the temporal domain}
The two input signals $X$ and $Y$ are modeled by bivariate random numbers with normal distribution, zero mean, standard deviation $\sigma = 1.0$ and correlation
coefficient $\rho = 0.0;0.05;0.1;0.2$.  Such $\rho$ are typical of many radio interferometric observations when weak signals are processed.
The number of samples in each data sequence is $n$ and the number of repetitions of the correlation is $m$.
Outliers $rfi_{i}$  added to the inputs $x_{i}$ and  $y_{i}$  are modeled by the following  expression:
\begin{equation}
rfi_{i}=s_{i}\times sign(z_{i})\times (A_{rfi}+\sigma _{rfi}\times u_{i}),
\end{equation}
where $s_{i}$ is the Poisson random value with the parameter $\lambda=0.01$, $A_{rfi}=20.0$ is the amplitude of  outliers randomized by
adding the normal random numbers $u_{i}$ with the standard deviation $\sigma _{rfi}=0.3A_{rfi}$.
The random polarities of the outliers are provided by the $sign(z_{i}), z_{i}$ is the other auxiliary random normal number.
Fig. 2  gives the  example of a sequence of  input samples.
  \begin{figure}
   \centering
   \includegraphics[width=8cm]{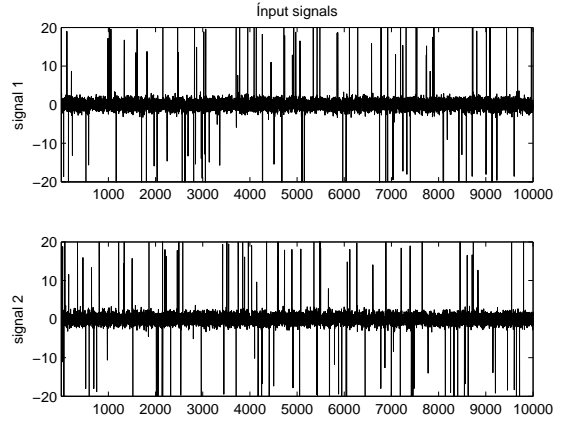}
      \caption{Input signals representing Gaussian noise + impulse-like outliers, the number of samples $n=10000$.}
         \label{FigWin2}
   \end{figure}

 The results of computer simulations, bias  and {\em rms} of the estimate of $\widehat{\rho}$ are presented in   Tables 1 - 5.
 The bias of the estimate $\widehat{\rho}$ is  $\rho: bias = \widehat{\rho}-\rho$.
 The confidence intervals are calculated for the  $95\%$-confidence level
 and for $rms_{\widehat{\rho}}=1/\sqrt{mn}$ for  $\widehat{\rho}$ and
  $rms_{\widehat{rms}}=1/\sqrt{2mn}$ for {\em rms} (the approximations that are valid for small $\rho$ and large $mn$).

\begin{table*}
   \caption[]{Bias  and
 {\em rms} of $\widehat{\rho }$   for the Spearman and Kendall correlators, $ n=1000, m=1000$, $95\%-$confidence intervals\\
  for bias: $\delta_{bias}=\pm0.00205$,
 for {\em rms}: $\delta_{rms}=\pm0.0014$. }
 \centering
\begin{tabular}{c|cc|cc|cc}
  \hline
  \multicolumn{7}{c}{ Without outliers }\\
  \hline

\multicolumn{7}{c}{\hspace{1.2cm}Pearson \hspace{1.8cm} Spearman  \hspace{1.8cm}   Kendall}\\
\hline
$\rho$ & Bias & Rms &  Bias & Rms &  Bias & Rms\\
\hline
 0.00&  0.000013& 0.03265& 0.000101   & 0.03425& 0.000100  & 0.03428 \\
 \hline
 0.05& -0.000672  & 0.03245  & -0.000436 & 0.03391 &-0.000366  & 0.03396 \\
 \hline
 0.10& 0.000075   &0.03268   & 0.000431 & 0.03454 & 0.000511& 0.03452\\
 \hline
 0.20 & 0.000445 & 0.03023  & 0.000696  & 0.03217 &0.000817 & 0.03207\\
  \hline
  \hline
    \multicolumn{7}{c}{ Without outliers }\\
   \hline
 \multicolumn{7}{c}{\hspace{1.2cm}Pearson \hspace{1.8cm} Spearman  \hspace{1.8cm}   Kendall}\\
   \hline
$\rho$ & Bias & Rms &  Bias & Rms &  Bias & Rms\\
\hline
 0.00&0.00156 & 0.03075& 0.000416 & 0.03453&0.000419&  0.03455 \\
\hline
 0.05&-0.04115    & 0.03026& -0.000179 & 0.03399& -0.000131&0.03405\\
 \hline
 0.10&  -0.07844 & 0.03212 & -0.003240  &0.03393 & -0.003159 &0.03392 \\
  \hline
  0.20 & -0.16020 & 0.03144&-0.007427 & 0.03233  &-0.007243&0.03222 \\

\end{tabular}
\end{table*}

\begin{table*}

    \caption[]{Bias  and
 {\em rms} of $\widehat{\rho }$   for trimming, $\gamma=0.1, n=1000, m=10000$, $95\%-$confidence intervals \\
 for bias: $\delta_{bias}=\pm0.00062$,
 for {\em rms}: $\delta_{rms}=\pm0.00044$. }
    \centering
    \label{TableTrim}
   \begin{tabular}{c|cc|cc|cc|cc}
    \hline

     \multicolumn{9}{c}{\hspace{1.0cm} Without outliers \hspace{4.5cm} With outliers}\\
     \hline

     \multicolumn{9}{l}{\hspace{2.0cm}  Pearson \hspace{2.0cm} Trimming \hspace{2.0cm} Pearson\hspace{2.2cm} Trimming}\\
    \hline
    $\rho$ & Bias & Rms &  Bias & Rms &  Bias & Rms&  Bias & Rms\\
     \hline
    0.00&  -0.000176 &  0.031209   &  -0.000171 &0.041631 &  -0.000048 & 0.030746  & -0.000193  & 0.042383\\
    \hline
    0.05& -0.000398 & 0.031813&  -0.000975 & 0.042481& -0.039886   & 0.031722  & 0.000503  & 0.043066  \\
    \hline
    0.10& 0.000326  & 0.031324    & 0.000575  & 0.041588 & -0.079740  & 0.031434    & -0.000943 & 0.042246 \\
    \hline
    0.20&-0.000556 & 0.030111 &0.000614& 0.040910 & -0.160044   & 0.032241  & -0.002102& 0.041670 \\
   \end{tabular}

\end{table*}

\begin{table*}

    \caption[]{Bias  and
 {\em rms} of $\widehat{\rho }$   for Winsorization, $\gamma=0.1, n=1000, m=10000$, $95\%-$confidence intervals \\
 for bias:
  $\delta_{bias}=\pm0.00062$,
 for {\em rms}: $\delta_{rms}=\pm0.00044$. }
    \centering
    \label{TableWin}
   \begin{tabular}{c|cc|cc|cc|cc}
    \hline

    \multicolumn{9}{c}{\hspace{1.0cm} Without outliers \hspace{4.5cm} With outliers}\\
     \hline
     \multicolumn{9}{l}{\hspace{2.0cm}  Pearson \hspace{1.8cm} Winsorization \hspace{1.7cm} Pearson\hspace{1.8cm} Winsorization}\\
    \hline
    $\rho$ & Bias & Rms &  Bias & Rms &  Bias & Rms&  Bias & Rms\\
     \hline
    0.00& 0.000032 & 0.031597  & 0.000046 & 0.037387 & -0.000692&  0.032661   & 0.000086  &0.038446  \\
    \hline
    0.05& -0.000025 & 0.031979 & -0.000105  & 0.037754& -0.040128  & 0.032028  & -0.000886  & 0.038748  \\
    \hline
    0.10& -0.000151  & 0.031276  & -0.000396 & 0.037174  & -0.080514  & 0.032396   &  -0.002059 &  0.038072\\
    \hline
    0.20& 0.000010  & 0.030453   & -0.000253 &0.036746&-0.160392 & 0.032851   & -0.002856  & 0.037700\\
   \end{tabular}

\end{table*}

\begin{table*}

    \caption[]{Bias  and
 {\em rms} of $\widehat{\rho }$   for median absolute deviation (MAD), $n=1000, m=10000$, $95\%-$confidence intervals\\
  for bias:
 $\delta_{bias}=\pm0.00097$,
 for {\em rms}: $\delta_{rms}=\pm0.00069$. }
    \centering
    \label{TableMAD}
   \begin{tabular}{c|cc|cc|cc|cc}
    \hline

    \multicolumn{9}{c}{\hspace{1.0cm} Without outliers \hspace{4.5cm} With outliers}\\
    \hline
     \multicolumn{9}{l}{\hspace{2.0cm}  Pearson \hspace{2.3cm} MAD \hspace{2.2cm} Pearson\hspace{2.3cm} MAD}\\
    \hline
    $\rho$ & Bias & Rms &  Bias & Rms &  Bias & Rms&  Bias & Rms\\
     \hline
    0.00& 0.000073 & 0.031411 &-0.000057& 0.051588& 0.000246& 0.031372 & -0.000097    &0.052656 \\
    \hline
    0.05&  0.000073  & 0.031761  & -0.000315  & 0.052502 & -0.039889  & 0.032401  & 0.001113    & 0.053639  \\
    \hline
    0.10& 0.000064   & 0.031470 & -0.000396  & 0.051782& -0.080238 & 0.031536  & 0.002077   & 0.052777 \\
     \hline
     0.20& -0.000510   & 0.030379 & -0.000158 & 0.051560 &-0.080546 & 0.033214  & 0.002244  & 0.052823\\

   \end{tabular}

\end{table*}

\begin{table*}

    \caption[]{Bias  and
 {\em rms} of $\widehat{\rho }$   for $Q2$-estimator, $n=1000, m=1000$,  $95\%-$confidence intervals \\
 for bias: $\delta_{bias}=\pm0.00205$,
 for {\em rms}: $\delta_{rms}=\pm0.0014$.}
    \centering
    \label{TableMAD}
   \begin{tabular}{c|cc|cc|cc|cc}
    \hline
    \multicolumn{9}{c}{\hspace{1.0cm} Without outliers \hspace{4.5cm} With outliers}\\
    \hline
     \multicolumn{9}{l}{\hspace{2.0cm}  Pearson \hspace{2.3cm} $Qn$ \hspace{2.5cm} Pearson\hspace{2.4cm} $Qn$}\\
    \hline
    $\rho$ & Bias & Rms &  Bias & Rms &  Bias & Rms&  Bias & Rms\\
     \hline
    0.00&  0.000488  & 0.031669 &0.000541& 0.035316  & -0.000122& 0.030729  & 0.000446   & 0.037311\\
    \hline
    0.05&  0.001529   & 0.031255 & 0.001392  & 0.033659 & -0.039906 & 0.032105& 0.000570   & 0.035250\\
    \hline
    0.10& 0.000362   & 0.031255   &  0.000277  & 0.034868 &-0.081281  & 0.031905   & 0.001994  & 0.037097\\
     \hline
     0.20& -0.001444    & 0.029626  & -0.001721& 0.032624&-0.158926 & 0.033120& 0.000594  & 0.034818\\

   \end{tabular}

\end{table*}

   \begin{figure}
   \centering
   \includegraphics[width=8cm]{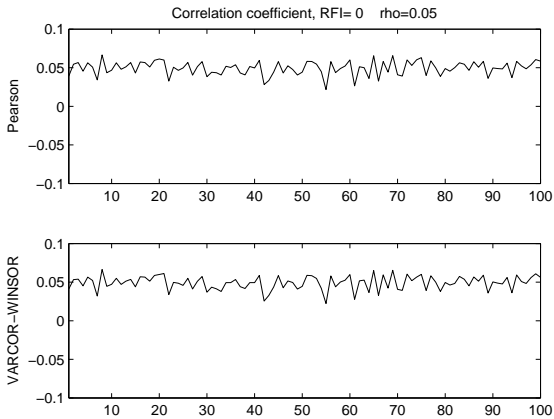}
      \caption{Correlation coefficients in the absence of outliers for Pearson's correlator  (1), (upper panel), and Winsorized correlator (7),
       (lower panel), each of the $m=100$  correlation coefficients are calculated for $n=1000$ data samples, $\rho=0.05$.}
         \label{FigWin1}
   \end{figure}

   \begin{figure}
   \centering
   \includegraphics[width=8cm]{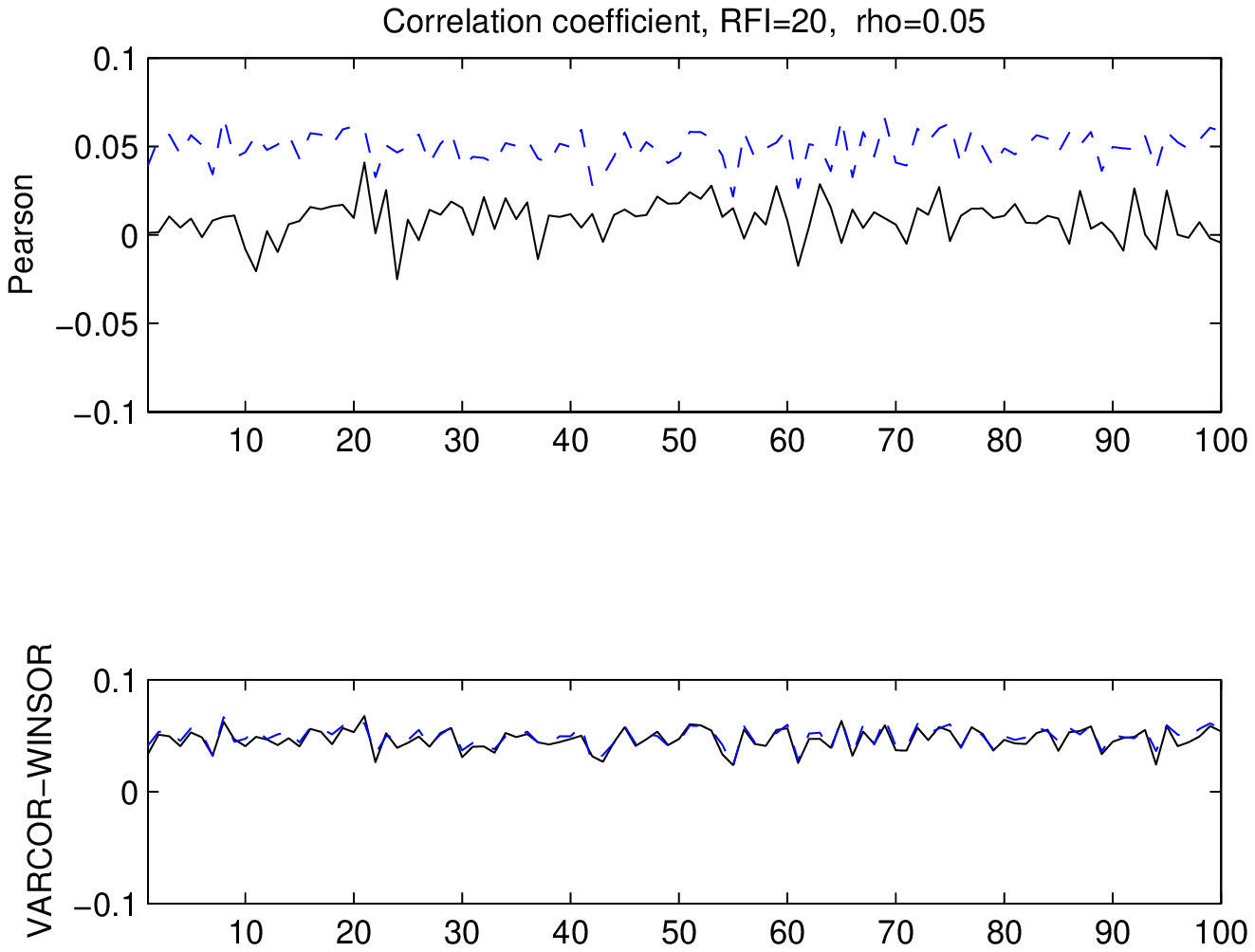}
      \caption{Correlation coefficients in the presence of outliers for Pearson's correlator (1), (upper panel), and Winsorized correlator (7),
      (lower panel), $A_{rfi}=20.0, \rho =0.05$.
      The solid line in the upper panel shows the total loss of correlation, the dashed line corresponds to the absence of outliers, while the solid and dashed lines in the lower panel practically coincide.}
         \label{FigWin3}
   \end{figure}

Looking at  Tables 1 - 5 several remarks can be made.\\

1. In the absence of outliers,  robust correlators reproduce practically the same values of $\widehat{\rho}$ as the Pearson's correlator (not showing
a significant bias) and keeping the effectiveness close to  $1/\sqrt {mn}$ (except MAD). Fig. 3  shows this for correlator (7) with Winsorization.\\
2. The output of the Pearson's correlator is dramaticaly  {\bf decorrelated}  with the growth of the outlier's
amplitudes whereas robust correlators analyze local statistics ($n$ samples) and eliminate outliers regardless of their amplitudes,
 see Fig. 4.\\
3. In the presence of outliers,   robust correlators  show a small positive bias which is equal to $\approx 1.0\%$ of $\rho$.\\
4. Spearman's and Kendall's correlators (Table. 1)  provide  a considerable reduction of bias in  the presence of RFI   compared to  the
reduction of bias in  Pearson's correlator and keep the
{\em rms}, i.e., there is no loss in effectiveness within the limits of error. These correlators require more operating capacity per lag,
due to the sorting and ranking of samples (this is also valid for the following algorithms).\\
5. Trimming and Winsorization (Tables 2 and 3) show good results with regard to bias and effectiveness. The {\em rms} for trimming is increased $\approx10\%$
than for Winsorization. The choice of $\gamma=0.1$  presumes a  percentage of outliers  of less than $10\%$. So there is a considerable
safety margin here but the growth of {\em rms}  for $\gamma=0.1$ is insignificant. In practice, an adaptive choice of $\gamma$ is possible in each
sub-band when
the  value of $\gamma$ is chosen after taking into account a  real RFI environment  situation, i. e., the  presence or absence of RFI and its duty cycle.\\
6. Median  absolute deviation (MAD) (Table. 4) gives the best results for the bias, but as  predicted theoretically, it  has the largest {\em rms}:
$\approx 1.65$ greater than $1/\sqrt {mn}$.\\
7. The $Qn$ estimate (Table. 5) requires more operating capacity than the others. The bias is as little as for the other algorithms, the {\em rms}  is slightly
larger ($\approx 10\%$) than $1/\sqrt {mn}$.\\
 8. The proposed methods are effective in the presence of strong impulse-like outliers.  When not intercepted, they decorrelate the correlator's output and it
is practically impossible to improve this situation with  subsequent processing.\\
Low-amplitude outliers are more difficult to detect because they are similar to the rare Gaussian noise
 spikes, but their decorrelation impact is also
much weaker. For example, with an RFI amplitude $\approx 3.0\sigma$, $\lambda=0.01$ and $\rho=0.1$ (signal-of-interest), the output of Pearson's correlator
is $0.092$ and the output of Spearman's correlator  $0.098$.\\
 9. The situation is different when outliers are correlated at the inputs of the correlator. In this case outliers produce an excessive, false  correlation even
in the absence of a  signal-of-interest, $\rho=0.0$. For example, for strong $100\%$-correlated   RFI with an amplitude equal to $30.0\sigma$, $\lambda=0.001$
and $\rho=0.0$,
the output of Pearson's correlator
is $0.349$  and   the output  of the MAD-correlator  is $0.001$.\\
 For correlated RFI, other methods
exploiting this correlation property
  can be  more effective, see (Ellingson\&Hampson\cite{ellingson}, Jeff et al.{\cite{jeff},
Kesteven\cite{kesteven}, Cornwell et al.\cite{corn}).

\subsection{Mitigation of outliers  in the frequency  domain}
Narrow-band RFI can be detected as an impulse-like outlier  in the frequency domain. The following example illustrates the application of trimming (8) in this case.
RFI is generated as a phase-modulated sinusoidal carrier, see Fig. 5. The power spectrum of the unmodulated carrier is shown in  the upper panel
and in the lower panel, the power spectrum  of the randomly phase-modulated signal  is represented:
\begin{equation}
s=A_{rfi}sin(2\pi Fi+(\pi/2)rect(i)), i=1...n
\end{equation}
where $rect(i)$ is a rectangular wave with  Poisson's law distributed random jumps from $-1$ to $+1$, and the parameter of Poisson  distribution $\lambda=0.01$.
 \begin{figure}
  \includegraphics[width=8cm]{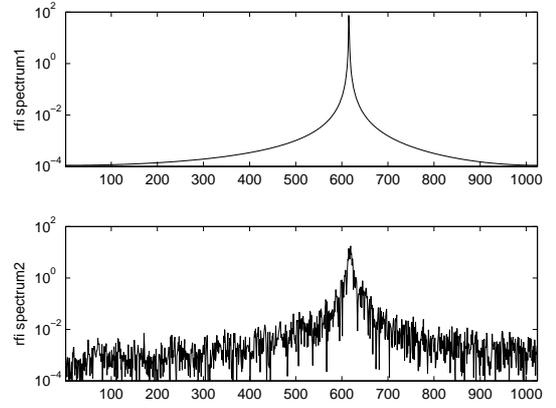}
  \caption{Power spectra of the unmodulated sinusoidal carrier (upper panel) and the randomly-phased modulated carrier with the phase jumps $\pm\pi/2$.}
 \end{figure}
The mixture of two input signals and their power spectra are represented in Fig. 6. The amplitudes $A1_{rfi}=A2_{rfi}=0.5$ and frequencies $F1=0.3$ and
$F2=0.4$. Therefore, RFI is not visible in the temporal domain but is easily visible in the spectral domain after  FFT with $n=2048$ as two bursts.
 \begin{figure}
  \includegraphics[width=8cm]{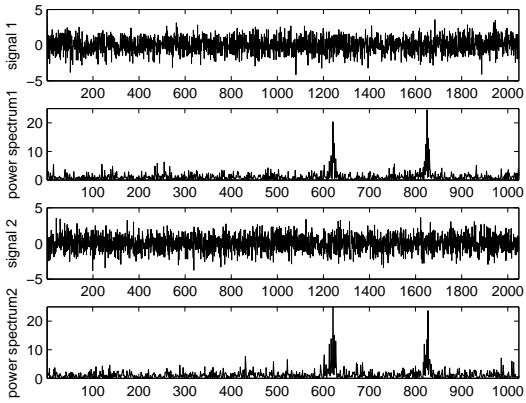}
  \caption{Input signal $x$  noise +RFI (upper panel) and its power spectra (upper middle panel), input signal $y$ (lower middle panel) and its power spectra
  (lower panel)}
 \end{figure}

The statistics of the power spectra are analyzed following (8) and the indexes of outliers exceeding the level equal to the  $1.5\%-$percentile are used to
reject the samples of the complex spectra of signals $\widetilde{x}$ and $\widetilde{y}$,  where $\widetilde{(.)}$ denotes the Fourier  transform.
These indexes are tagged with
the indicator function $ind(i)$ taking the numbers 0 or 1,  $i=1..n$ and assigned to the sequence of n samples of the complex spectra
 $\widetilde{x}$ and $\widetilde{y}$.
The ``sum-difference'' correlator (7) is used  in the spectral domain (FX-correlator):
\begin{eqnarray}
var(sum)=var(\Re(sum))+var(\Im(sum)), sum=\widetilde{x}+\widetilde{y},\nonumber\\
var(dif)=var(\Re(dif))+var(\Im(dif)), dif=\widetilde{x}-\widetilde{y},
\end{eqnarray}
Robust variance using the trimming algorithm (8) is applied while calculating (15): outliers are marked by the indicator function $ind(i)$.
The results of computer simulation for $n=2048$ and $m=100$ are shown in Fig. 7, 8 and 9. Fig.  7 shows the $m$ outputs of the Pearson's correlator, upper panel,
and the robust correlator (15), lower panel, for  the correlation coefficient
of the input signals $\rho=0$, i.e., for the uncorrelated inputs, except  RFI, which are $100\%$ correlated. The upper panel shows considerable
bias, while the fluctuations of correlator output in the lower panel vary around zero. Fig. 8 gives the same  situation but for $\rho=0.1$ and Fig. 9  - for
$\rho=0.2$. In all these examples a considerable bias is visible for the Pearson's correlator and there is an absence of  bias for
the ``sum-difference'' correlator.
\begin{figure}
  \includegraphics[width=8cm]{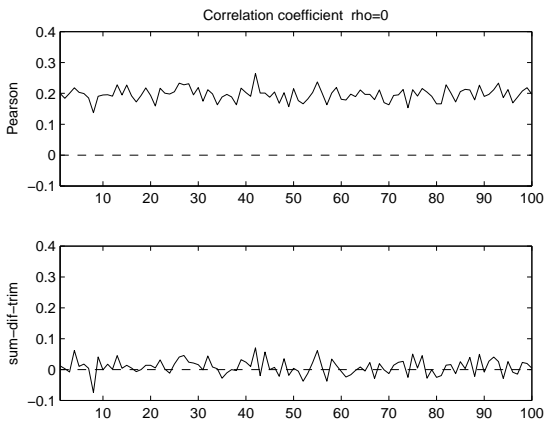}
  \caption{Pearson's correlator output (upper panel) and the ``sum-difference'' correlator ouput
  (lower panel), $\rho=0.0$.}
 \end{figure}
 \begin{figure}
  \includegraphics[width=8cm]{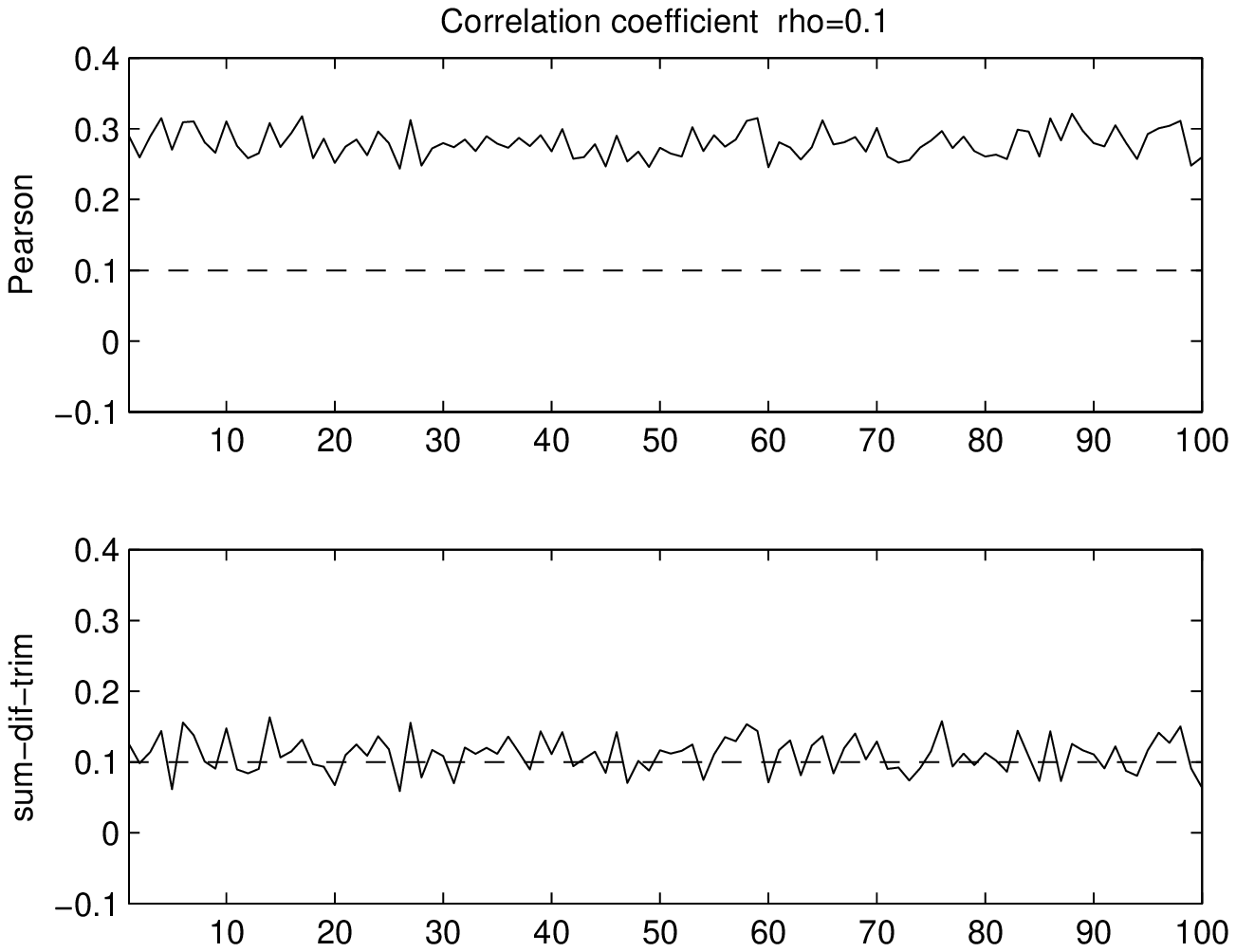}
  \caption{Pearson's correlator output (upper panel) and the ``sum-difference'' correlator ouput
  (lower panel), $\rho=0.1$.}
 \end{figure}
 \begin{figure}
  \includegraphics[width=8cm]{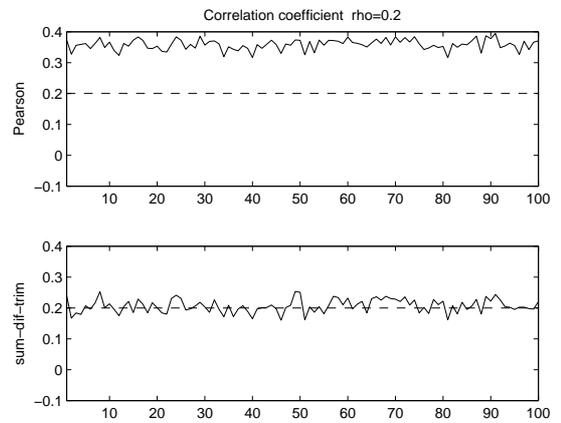}
  \caption{Pearson's correlator output (upper panel) and the ``sum-difference'' correlator ouput
  (lower panel), $\rho=0.2$.}
 \end{figure}

 Of course, other post-correlation methods can be  used in the case of  narrow-band persistent RFI with a stable  spatial orientation,
  for example, (Cornwell et al.\cite{corn}). But in the case of
 sporadic burst-like RFI, the proposed pre-correlation  statistical analysis is more appropriate: only $n\approx2000$ samples were used for each
 correlator input, which corresponds to a  microsecond time scale for typical bandwidths.

 \subsubsection{Processing of observational data}
 Several examples of applications of the aforementioned  algorithms are presented here.\\
  The auto-spectrum in Fig. 10 is calculated using the ``raw'' data recorded at LOFAR CS1 for three hours. Data consisting of  complex eight-bit   numbers
 and having a  sample  time interval equal to 6.4 $\mu sec$ were recorded on
 the subband with  central frequency
 $f_{0}=205.781MHz$, bandwidth $\Delta f=156.25KHz$. This time-frequency presentation  corresponds to 32 frequency channels, i.e.,
 the frequency resolution is $\delta f\approx 4.88KHz$, and the integration time $t\approx 1.7s$. \\
 Fig. 11 demonstrates the auto-spectrum calculated from the same data  for 1024 frequency channels. The high spectral
 resolution, $\delta f\approx 153Hz$, permits the separation of RFI spikes, i.e., not  smoothing them, as would have been
 in the case of 32 channels, see Fig. 10. Only half of the 3-hour duration auto-spectrum is shown in Fig. 11
  (due to the constraints of computer memory). The auto-spectrum in Fig. 11 is calculated without  any censoring of outliers. \\
  Fig. 12  shows the ``cleaned''  auto-spectrum where  Winsorization was
 applied.
 The Winsorized spectrum in Fig. 12 is smoothed so as to get a final frequency resolution  $\delta f\approx 4.88KHz$ as in Fig. 10.\\

 \newpage
\suppressfloats
  \begin{figure}[h]
  \includegraphics[width= 7.5cm]{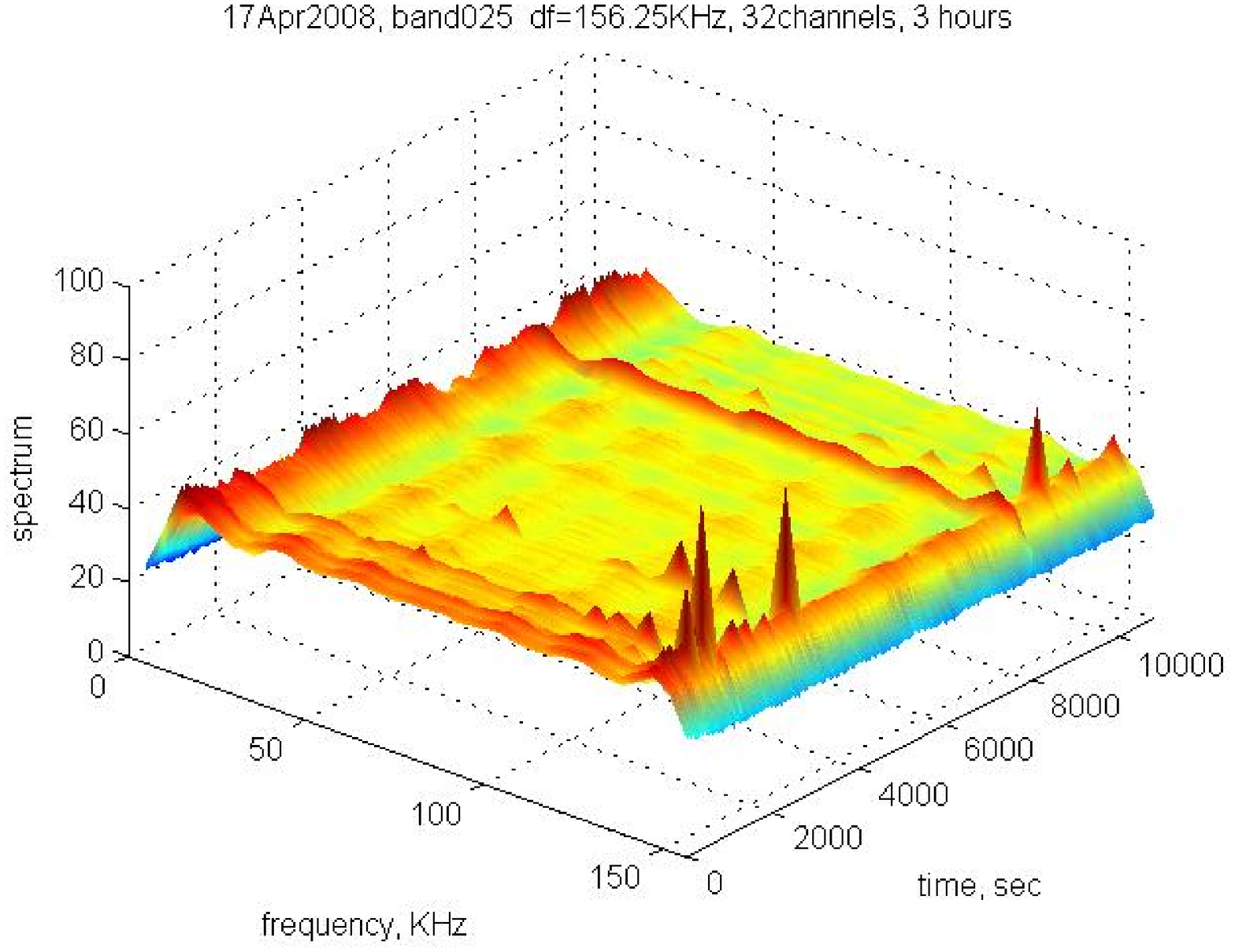}
 \caption{Auto-spectrum of  system noise +RFI from LOFAR CS1 at the central  frequency $f_{0}=205.781MHz$, bandwidth $\Delta f=156.25KHz$,
 frequency resolution $\delta f\approx 4.88KHz$, integration time $t\approx 1.7s$.}
  \end{figure}
  \begin{figure}[h]
  \includegraphics[width=7.5cm]{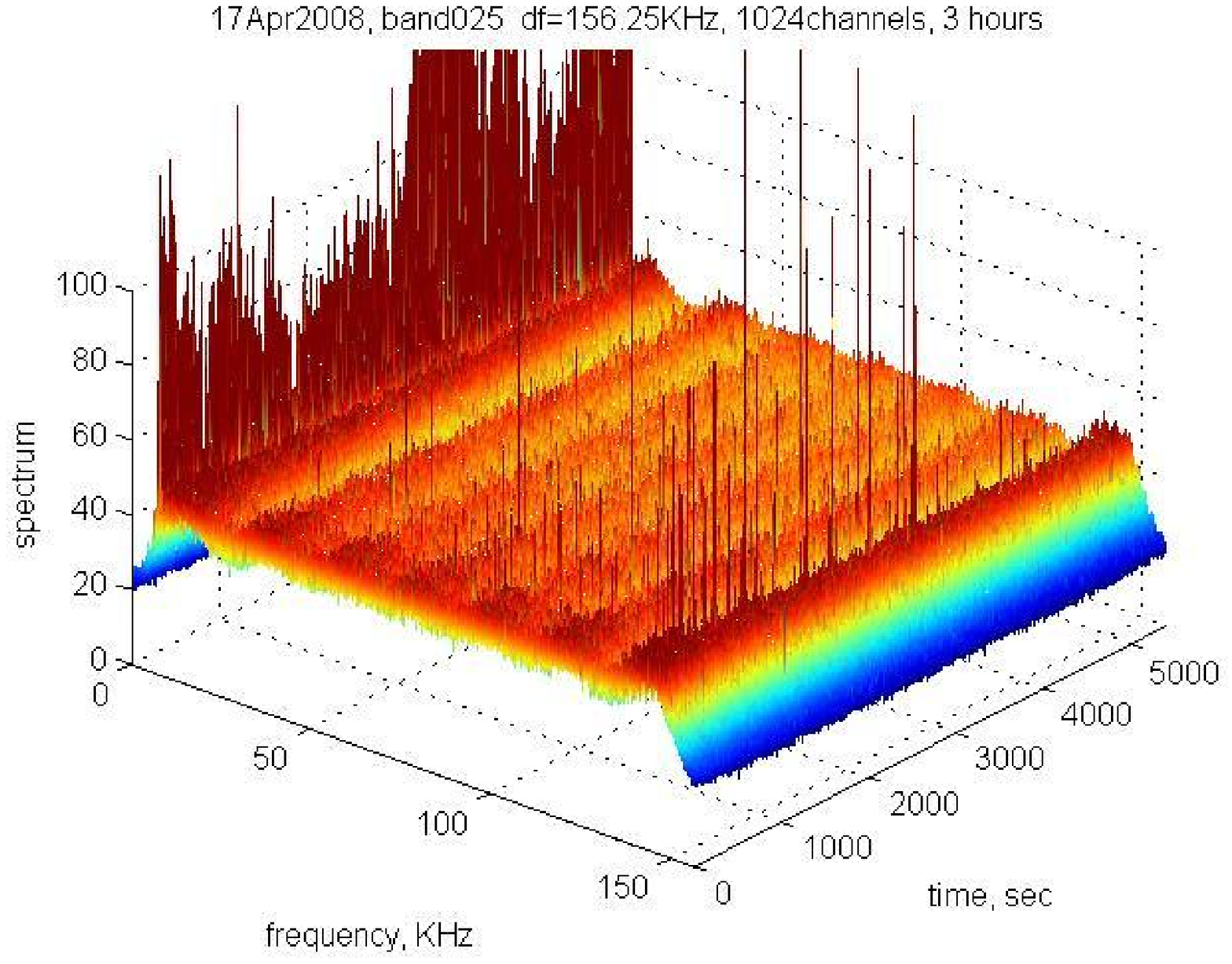}
 \caption{Auto-spectrum of  system noise +RFI from LOFAR CS1 at the central  frequency $f_{0}=205.781MHz$, bandwidth $\Delta f=156.25KHz$,
 frequency resolution $\delta f\approx 153Hz$, integration time $t\approx 1.7s$.}
  \end{figure}
  \begin{figure}[h]
  \includegraphics[width=7.5cm]{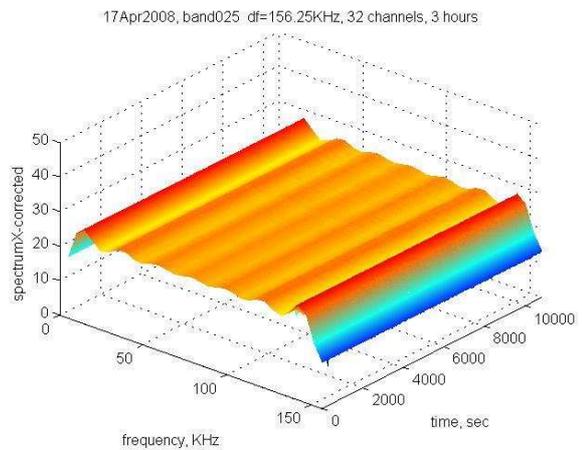}
 \caption{Auto-spectrum of the same data as in Fig. 11  calculated with Winsorization.}
  \end{figure}

   \begin{figure}[h]
  \includegraphics[width=7.5cm]{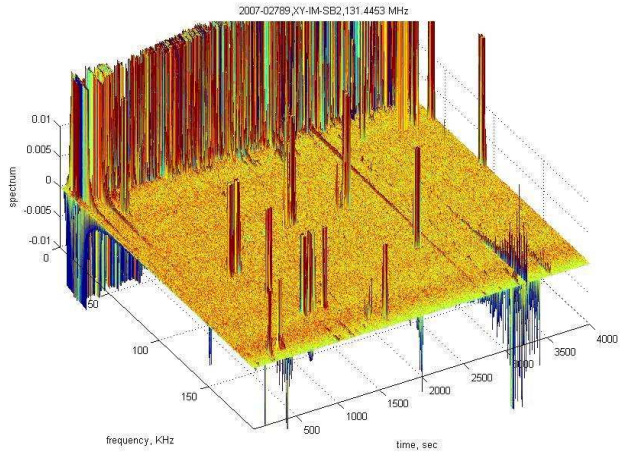}
  \caption{Post-correlation cross-spectrum, central  frequency $f_{0}=131.4453MHz$, bandwidth $\Delta f= 156.25KHz$, frequency resolution $\delta f=610Hz$,
  time resolution $\delta t=1sec$.}
 \end{figure}
   \begin{figure}[h]
  \includegraphics[width=7.5cm]{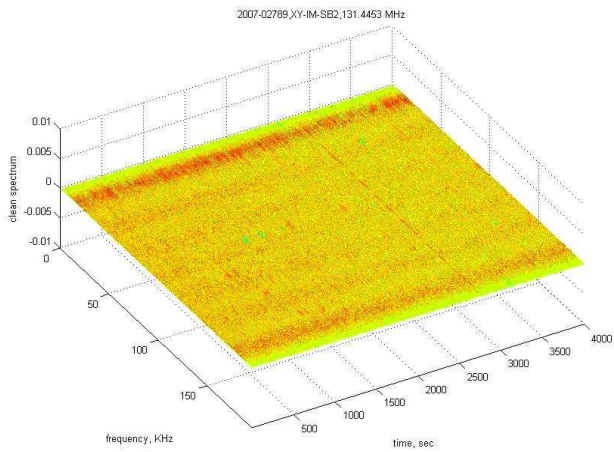}
  \caption{Post-correlation cross-spectrum from Fig. 13  with  outliers removed.}
 \end{figure}
   \begin{figure}[h]
  \includegraphics[width=7.5cm]{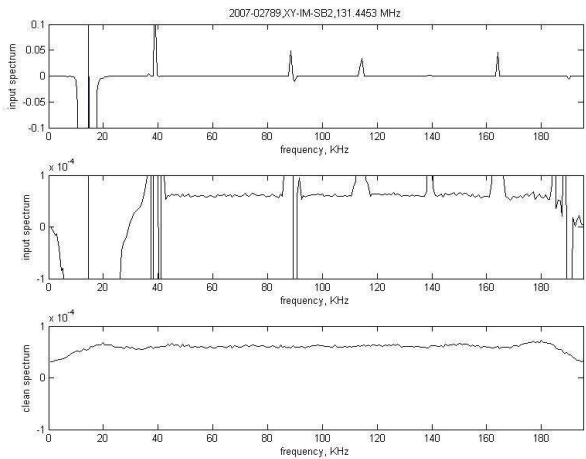}
  \caption{Averaged cross-spectra corresponding to Fig. 13  (upper  and middle (zoomed) panels) and to Fig.  14 (lower panel).}
 \end{figure}

 \clearpage
 This example shows the importance
 of a sufficiently  high frequency resolution.
All RFI visible  in Fig. 11  are  invisible in the temporal domain; they are too weak
 and they are ``below" the system noise. The low level of RFI also made it  necessary to average the instantaneous power spectra obtained after each FFT.
 The number of  averaged spectra is 256 and thus the time resolution in the Fig. 11  is $6.4\times 10^{-6}\times 1024\times256=1.677sec$.
 The  ripple structure  clearly visible in the auto-spectra is produced by the transfer functions of  LOFAR digital filters separating the whole
 input bandwidth into dozens of sub-bands with the partial bandwidths $\Delta f=156.25KHz$ (or $200KHz$).

 The censoring of outliers may also be  useful in the case of  post-correlation data produced by the LOFAR correlator. The filter bank of the LOFAR backend
 divides each of the 156KHz-bandwidth sub-bands into  256 narrow ``sub-sub-bands"  with a bandwidth  equal to $610.3516Hz$.
  The sample time interval after preliminary averaging is 1 sec. This good time-frequency resolution allows  the fine-grain structure of RFI to be seen.

  The following example in Fig. 13 shows  the post-correlation {\bf cross-spectrum} on frequency $f_{0}=131.4453MHz$.\\
  Fig. 14  shows the corresponding Winsorized cross-spectrum and Fig. 15   shows the averaged cross-spectra along
  the whole time interval $4\times10^{4}sec$ - without Winsorization (upper and middle panels) and with Winsorization (lower panel).
   The weak correlated component (the signal-of-interest) is produced by
   some cross-polarization effects and bears a ripple structure similar to the auto-spectra.\\

\section{Conclusions}

   \begin{enumerate}
      \item  Estimates of the correlation coefficient  are an  important
part of both  classical  and of robust  statistics. Statistical analysis of raw data with the finest available time
and frequency resolution can help during observations in an RFI
contaminated environment. Growing concern
about RFI pollution  should persuade  the  radio astronomy community
to pay more attention to the variety of algorithms developed in the
realm of robust  statistics.  This framework  of robust estimates puts
the successfully tested  blanking of RFI on a more stable
foundation.
      \item Statistically faithful, robust  estimates of the  correlation coefficient are
especially appropriate for application in an  impulse-like strong
RFI  environment. RFI is effectively  suppressed and the
accompanying  bias and effectiveness are  tolerable.  The aforementioned
robust  algorithms can be  usefully applied in these particular situations.
      \item The choice of a particular algorithm depends upon the type and
intensity of the RFI. The proportion of RFI  presence in data is also
important. The type of implementation may determine the choice:
{\it off-line}  or {\it real-time}. It is difficult to
equip existing conventional hardware correlators   with these tools. Future generations of radio  telescopes
(LOFAR, ATA, SKA)  will generate such huge amounts of data
that  {\it real-time} processing is  vital: DSP, FPGA or
supercomputers are possible solutions. Software correlators are especially suited to the implementation
of  robust schemes as optional subroutines.
   \end{enumerate}

\begin{acknowledgements}
    My discussions with Ger de Bruyn and Jan Noordam have been very
     helpful and I am  grateful for their continued  support.
\end{acknowledgements}

\end{document}